\documentclass[11pt]{article}
\usepackage[utf8]{inputenc}
\usepackage{bm}
\usepackage{amsmath}
\usepackage{graphicx}
\usepackage{geometry}
\begin{document}
\title{Non-intuitive Computational Optimization of Illumination Patterns for Maximum Optical Force and Torque}
\author{Yoonkyung E. Lee,\textsuperscript{1} Owen D. Miller,\textsuperscript{2,3} M. T. Homer Reid,\textsuperscript{2} Steven G. Johnson,\textsuperscript{2} \\ and Nicholas X. Fang\textsuperscript{1}}
\date{\vspace{-5ex}}
\maketitle 

\bigskip

\begin{small}
\noindent \textsuperscript{1}Department of Mechanical Engineering, Massachusetts Institute of Technology, Cambridge, Massachusetts 02139, USA\\
\textsuperscript{2}Department of Mathematics, Massachusetts Institute of Technology, Cambridge, Massachusetts 02139, USA\\
\textsuperscript{2}Currently with the Department of Applied Physics, Yale University, New Haven, Connecticut 06520, USA
\end{small}
\medskip

\noindent \textsuperscript{*}nicfang@mit.edu

\begin{abstract}
This paper aims to maximize optical force and torque on arbitrary
micro- and nano-scale objects using numerically optimized structured
illumination. By developing a numerical framework for computer-automated
design of 3d vector-field illumination, we demonstrate a 20-fold enhancement
in optical torque per intensity over circularly polarized plane wave
on a model plasmonic particle. The nonconvex optimization is efficiently
performed by combining a compact cylindrical Bessel basis representation
with a fast boundary element method and a standard derivative-free,
local optimization algorithm. We analyze the optimization results
for 2000 random initial configurations, discuss the tradeoff between
robustness and enhancement, and compare the different effects of multipolar
plasmon resonances on enhancing force and torque. All results are
obtained using open-source computational software available online.
\end{abstract}

\bigskip

\noindent OSCIS code (350.4855) Optical Tweezers or optical manipulation;(140.7010); (090.1760) Computer holography; Diffraction and Gratings: Optical Vortices;  (250.5403) Plasmonics; (290.2200) Extinction

\section{Introduction}

We show how large-scale computational optimization \cite{bertsekas_nonlinear_1999,powell_fast_1978,johnson_nlopt_2014}
can be used to design superior and non-intuitive structured illumination
patterns that achieve 20-fold enhancements (for fixed incident-field
intensity) of the optical torque on sub-micron particles, demonstrating
the utility of an optimal design approach for the many nanoscience
applications that rely on optical actuation of nanoparticles \cite{grier_revolution_2003,dholakia_optical_2008,agarwal_manipulation_2005}. 

Recent advances in nanoparticle engineering \cite{kelly_optical_2003,xia_shape-controlled_2005}
and holographic beam-generation via spatial light modulators (SLMs)
\cite{heckenberg_generation_1992,curtis_dynamic_2002,di_leonardo_computer_2007,chen_generation_2011}
and other phase-manipulation techniques\cite{karimi_efficient_2009,dolev_surface-plasmon_2012,schulz_integrated_2013,chen_creating_2015}
have created many new degrees of freedom for engineering light–particle
interactions beyond traditional optical tweezers. Enhanced and unusual
optical forces and torques can be engineered by designing material
objects \cite{liu_radiation_2005,liu_light-driven_2010,lehmuskero_ultrafast_2013,arita_laser-induced_2013,chen_optical_2016}
and/or structured illumination, with the latter including ``tractor
beams'' \cite{novitsky_single_2011,sukhov_negative_2011} and beams
carrying optical angular momentum \cite{simpson_mechanical_1997,torres_twisted_2011,chen_negative_2014,lehmuskero_plasmonic_2014}.
These increased degrees of freedom pose an interesting design challenge:
for a given target object, what is the optimal illumination pattern
to produce the strongest optical force or torque? While a small number
of Gaussian beam parameters can be manually calibrated for optimal
performance using manual trial-and-error \cite{singer_three-dimensional_2000},
an arbitrary 3D vector field requires a more targeted approach. Moreover,
optimization of a 3d vector field is highly nonconvex by nature, and
posesses many local optima due to wave interference and resonance.
When exploring so many parameters, a large number of scattering problems
must be solved efficiently, which requires careful design of the optimization
framework. 

Research in computational optimization of optical actuation has focused
on the design of new material geometries \cite{gersborg_maximizing_2011,hajizadeh_optimized_2010}
and on the improvement of multiplexed optical traps for microscale
dielectric particles (holographic optical tweezers) \cite{tolic-norrelykke_matlab_2004,polin_optimized_2005,martin-badosa_design_2007,bianchi_real-time_2010,cizmar_holographic_2010,tao_tao_3d_2011,lapointe_towards_2011}.
However, no computational method has been available to design structured
illumination for unconventional target objects that are nonspherical,
lossy, or nanometer-scale, thereby requiring a costly full-wave numerical
simulation for computing optical force and torque.

We present a compact and rapid computational framework to optimize
structured illumination for the mechanical actuation of an arbitrary
target object. We combine (i) a compact Bessel-basis representation
(Sec.\ref{subsec:Analytical-Representation-of}); (ii) a numerical
solver based on boundary element method (BEM) that discretizes the
surfaces of a 3d scattering problem to form a BEM matrix, and solves
hundreds of thousands of incident-field configurations using the same
matrix (Sec.\ref{subsec:Numerical-Solver}); (iii) an appropriate
optimization algorithm that exploits the smoothness of the nonconvex
and nonlinear optimization problem (Sec.\ref{subsec:Optimization-Algorithm});
and (iv) a suitable figure of merit (FOM) and optimization constraints
(Sec.\ref{subsec:Figure-of-Merit}). As a result, we rapidly attain
many-fold improvements in optical forces and torques over random field
or plane wave illuminations (Sec.\ref{subsec:Optimization-Results}),
and discuss the tradeoff between enhancement and robustness of optimization
(Sec.\ref{subsec:Robustness}). Furthermore, a given material object
may have scattering resonances at various frequencies, and the choice
of frequency for the incident field has several important implications.
When comparing interactions with two different resonances, we were
able to distinguish the impact of the change in the resonant field
pattern from the change in the resonance lifetime. Controlling for
the change in lifetime, we found that torque seems to favor higher-order
(\emph{e.g.} quadrupole) resonances with greater angular momentum,
while force seems to favor lower-order (\emph{e.g.} dipole) resonance
with greater field intensity within the particle (Sec.\ref{subsec:Force}).

\section{Optimization Framework\label{sec:Optimization-Framework}}

\begin{figure}
\begin{centering}
\includegraphics[clip,height=7cm]{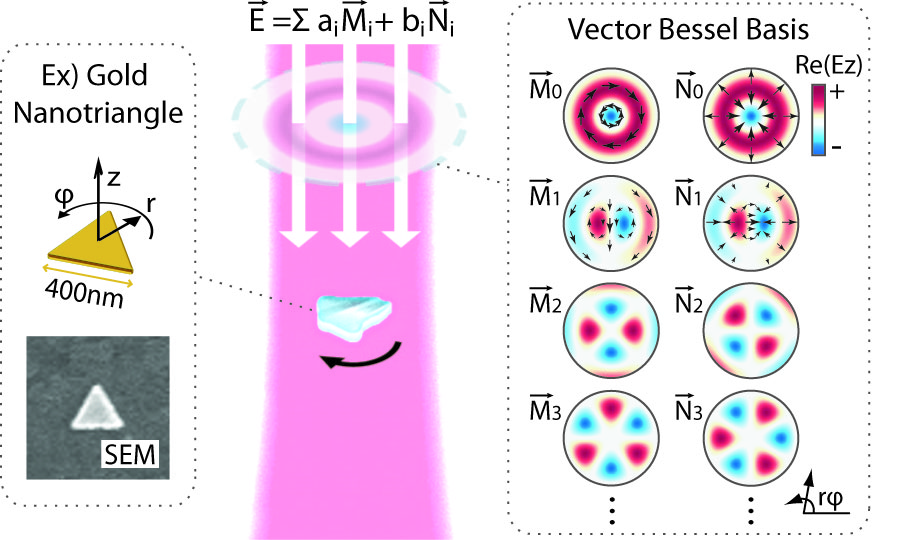}
\par\end{centering}
\centering{}\caption{Schematic of a structured vector-field illumination, analytically
represented with a vector Bessel basis. The right inset shows the
distributions for electric field (color) and polarization (black arrows)
of the vector Bessel basis. The illumination can be optimized to produce
maximum mechanical force and torque on an example target particle.
The gold nanotriangle in the left inset has edge length 400nm, thickness
40nm, and rounding diameter 30nm. The scanning electron microscope
(SEM) image shows an experimental sample fabricated using electron-beam
lithograpy. \label{fig:Schematic}}
\end{figure}

A structured-illumination optimization aims to find the best 3d vector
field, i.e., the one that maximizes a desired FOM for a given scattering
problem (see Fig.\ref{fig:Schematic}). Our choice of the Bessel
basis expansion is described in Sec.\ref{subsec:Analytical-Representation-of}.
Sec.\ref{subsec:Numerical-Solver} and Sec.\ref{subsec:Optimization-Algorithm}
discuss the BEM solver and the optimization algorithm. Lastly, Sec.
\ref{subsec:Figure-of-Merit} explains force and torque FOMs and the
corresponding optimization constraints. The entire numerical framework
is implemented with \texttt{C++}.

\subsection{Analytical Representation of Structured Illumination\label{subsec:Analytical-Representation-of} }

The computational design of structured illumination requires a compact
analytical representation of an arbitrary 3d vector field. The vector
field will contain spatial variations in both intensity and phase,
and must satisfy the vector wave equation \cite{j.d._jackson_classical_1962}.
The electric field $\mathbf{E}$ can be represented using a basis
expansion $\mathbf{E}=\sum_{i=0}^{N}c_{i}\boldsymbol{\phi}_{i}$,
where the complex scalar coefficient $c_{i}$ determines the relative
intensity and phase of each mode $\boldsymbol{\phi}_{i}$. 

The choice of coordinates and the basis functions \textbf{$\boldsymbol{\phi}_{i}$}
depends on the problem geometry. While the spherical coordinate system
is a common choice in Mie scattering \cite{bruning_multiple_1971,bohren_absorption_2004},
it requires a very large number of modes $N$ to represent light propagating
along a linear axis and potentially interacting with flat substrates
or SLMs. The cartesian coordinate system is also ill-suited because
it requires large $N$ to describe laser beams with a finite radius.
We find that the cylindrical coordinate system \cite{stratton_electromagnetic_2007,zhan_cylindrical_2009}
is well suited, requiring a small number of modes to describe structured
illumination with varying distributions of linear and angular momentum. 

Among the wide menu of cylindrical basis functions (e.g., Bessel,
Laguerre-Gaussian, Hermite-Gaussian, and so forth) \cite{zhan_cylindrical_2009},
we choose the Bessel basis \cite{rosen_pseudo-nondiffracting_1995,volke-sepulveda_orbital_2002}
for its compact analytical expression, derived from the scalar generating
function
\begin{equation}
\psi_{m}(r,\varphi,z)=J_{m}(k_{t}r)\exp(im\varphi+ik_{z}z),\label{eq:besspsi}
\end{equation}
where $J_{m}$ is the mth-order Bessel function, $k_{t}$ is the transverse
wavevector in $\hat{r}$, and $k_{z}$ is the longitudinal wavevector
in $\hat{z}$, satisfying $k_{t}^{2}+k_{z}^{2}=2\pi/\lambda$. The
ratio between $k_{t}$ and $k_{z}$ specificies the numerical aperture
of the basis, $\textit{NA}=\tan^{-1}(k_{t}/k_{z})$. Higher \emph{NA}
represents greater transverse momentum that can increase optical torque.
But the permissible range of \emph{NA} is often dictated by experimental
considerations, and the range of \emph{NA} in the optimization can
be set accordingly. 

Taking spatial derivaties of Eq.(\ref{eq:besspsi}) gives
\begin{align}
\mathbf{\mathbf{M}}_{i} & =\mathbf{\nabla}\times(\psi_{i}\mathbf{u}_{z}),\text{\qquad(azimuthal polarization)}\label{eq:Mi}\\
\mathbf{N}_{i} & =\frac{1}{k}\mathbf{\nabla}\times\mathbf{M}_{i}\text{,\qquad\ \;\;(radial polarization)}\label{eq:Ni}
\end{align}
where $\mathbf{u}_{z}$ is the unit vector in $\hat{z}$, and $\mathbf{\mathbf{M}}_{i}$
and $\mathbf{N}_{i}$ are the $i$th bases for azimuthal and radial
polarizations, respectively (right inset of Fig.\ref{fig:Schematic}).
The incident electric field $\mathbf{E}_{\text{inc}}$ can be expressed
as: 
\begin{equation}
\mathbf{E}_{\text{inc}}(r,\phi,z)=\sum_{i=0}^{N}a_{i}\mathbf{M}_{i}+b_{i}\mathbf{N}_{i},\label{eq:Eexpand}
\end{equation}
where $a_{i}$ and $b_{i}$ are the complex scalar coefficients. The
Bessel basis produces the most compact expressions for $\mathbf{\mathbf{M}}_{i}$
and $\mathbf{N}_{i}$ because the magnitude of $\psi$ does not vary
with $z$, reducing $\partial/\partial z$ terms in Eqs.(\ref{eq:Mi},\ref{eq:Ni}).

Note that we intentionally decouple our optimization framework from
the idiosyncratic differences in the spatial resolution of the SLMs.
A wide variety of experimental methods (e.g., superposed pitch-fork
holograms) \cite{heckenberg_generation_1992} can be used to generate
beams expressed as Eq.(\ref{eq:Eexpand}), for a finite $N$ and \emph{NA}.
In this paper, we consider numerical apertures with opening angles
$\leq10^{\circ}$ and 12 basis functions ($N=5$).

\subsection{Numerical Solver\label{subsec:Numerical-Solver}}

The optimization process itself has no restrictions on the choice
of the numerical solver, so the biggest consideration is the computational
cost: the smaller the the better. We choose the Boundary Element Method
(BEM) \cite{chew_integral_2008,harrington_field_1996} for several
reasons. In comparison to other scattering methodologies such as the
finite-difference or finite-element methods, BEM is particularly well-suited
to the type of large-scale optimization problem requiring a rapid
update of the incident field for a given geometry and a given wavelength
$\lambda_{\text{opt}}$.
\begin{equation}
\underbrace{\boldsymbol{M}}_{\begin{smallmatrix}\text{fixed}\\
\text{BEM matrix}
\end{smallmatrix}}\underbrace{\mathbf{\mathbf{c}}}_{\begin{smallmatrix}\text{output}\\
\text{current}
\end{smallmatrix}}=\underbrace{\mathbf{f}}_{\begin{smallmatrix}\text{rapidly updated}\\
\text{input field}
\end{smallmatrix}}\label{eq:BEM}
\end{equation}
In Eq.(\ref{eq:BEM}), the BEM matrix $\boldsymbol{M}$ remains fixed for
a given geometry and frequency, while the column $\mathbf{f}$ representing
the incident field is rapidly updated at each step of the optimization
process. This allows hundreds of thousands of scattering configurations
to be computed on the order of a few hours. In addition, BEM projects
the 3d scattering problem onto a 2d surface mesh, thereby reducing
the computation volume by a factor of 1/2000 in the nanoparticle scattering
problem we consider. Lastly, recent improvements \cite{reid_efficient_2013}
have significantly increased the speed with which optical froce and
torque can be computed in BEM.

\subsection{Optimization Algorithm\label{subsec:Optimization-Algorithm}}

Structured-illumination optimization is nonlinear and nonconvex, such
that searching for a global optimum is prohibitively expensive. Therefore
we choose a local algorithm with random starting points. We choose
one of the simplest solutions available: constrained optimization
by linear approximation (COBLYA) \cite{powell_fast_1978}, a derivative-free
algorithm that exploits the smoothness of the problem. An open-source
implementation of COBYLA is available through NLopt \cite{johnson_nlopt_2014}.

\subsection{Figure of Merit and Optimization Constraints for Optical Force and
Torque\label{subsec:Figure-of-Merit}}

We consider two types of optical actuation with respect to the object
coordinate (left inset of Fig.\ref{fig:Schematic}); the force $F_{z}$
and torque $T_{z}$. In order to avoid the optimizer from increasing
the brightness of the beam indefinitely, we choose to divide the force
and torque by the average incident-field intensity on the particle
surface ($I_{\text{avg}}=\overline{\vert E_{\text{inc}}\vert^{2}}/2Z_{0}$,
where $Z_{0}$ is the impedance of free space), which is easily computed
in BEM. We choose the incident-field rather than the total-field intensity
to avoid penalizing high extinction efficiency. $I_{\text{avg}}$
is measured on the particle surface because we want to account for
the portion of the beam that interacts with the target particle, rather
than the entire beam. We choose nondimensionalized figures of merit:
\begin{eqnarray}
\text{FOM}_{F} & = & \frac{F_{z}}{I_{\text{avg}}}\cdot\left(\frac{\pi c}{3\lambda^{2}}\right),\label{eq:forceFOM}\\
\text{FOM}_{T} & = & \frac{T_{z}}{I_{\text{avg}}}\cdot\left(\frac{4\pi^{2}c}{3\lambda^{3}}\right),\label{eq:torqueFOM}
\end{eqnarray}
where the constants in parentheses reflect ideal single-channel scattering.
The largest \cite{hamam_coupled-mode_2007,kwon_optimal_2009,liberal_least_2014}
scattering cross-section into a single (spherical harmonic) channel
is $3\lambda^{2}/2\pi$, which when multiplied by single-photon changes
in linear ($2\hbar k$) and angular ($\hbar$) momentum per photon,
divided by the photon energy ($\hbar\omega$), yields the constants
in Eqs.(\ref{eq:forceFOM},\ref{eq:torqueFOM}). 

Optimization constraints can be added to suppress actuation in undesired
directions. We suppress actuations in directions other than $\hat{z}$
using smooth constraints: $\left(\vert\mathbf{F}\vert^{2}-F_{z}^{2}\right)/\vert\mathbf{F}\vert^{2}\leq0.01$
and $\left(\vert\mathbf{T}\vert^{2}-T_{z}^{2}\right)/\vert\mathbf{T}\vert^{2}\leq0.01$,
where the limiting value 0.01 is set to ensure that $F_{z}$ and $T_{z}$
exceed $99\%$ of the force and torque magnitudes $\vert\mathbf{F}\vert$
and $\vert\mathbf{T}\vert$, respectively. 

\section{Results and Discussion\label{sec:Force-and-Torque} }

We demonstrate our illumination-field optimization framework on the
gold nanotriangle illustrated in Fig.\ref{fig:Schematic}. Our previous
work \cite{lee_optical_2014} analyzes the optical force and torque
on such a particle for circularly-polarized (CP) planewave illumination.
CP planewave is a common incident-field choice \cite{marston_radiation_1984,friese_optical_1998,lehmuskero_ultrafast_2013,arita_laser-induced_2013}
for torque generation due to its intrinsic spin angular momentum,
but we find in our computational optimization that highly optimized
field patterns can show 20x improvement of $\text{FOM}_{T}$. The
wavelength of illumination in each optimization, $\lambda_{\text{opt}}$,
is chosen to correspond to the plasmonic resonance wavelengths of
the model particle. 

Sec.\ref{subsec:Optimization-Results} presents the distribution
of 2000 local-optimization results and discusses the optimized field-patterns.
Sec.\ref{subsec:Force} analyzes the wavelength-dependence of optical
force and torque for optimized illuminations, based on the choice
of $\lambda_{\text{opt}}$, and compares the results with the reference
force and torque from CP planewave. Lastly, the tradeoff between robustness
and enhancement is discussed in Sec.\ref{subsec:Robustness}.

\subsection{Illumination-field Optimization from 2000 Randomly Selected Initial
Configurations\label{subsec:Optimization-Results}}

\begin{figure}
\centering{}\includegraphics[height=6cm]{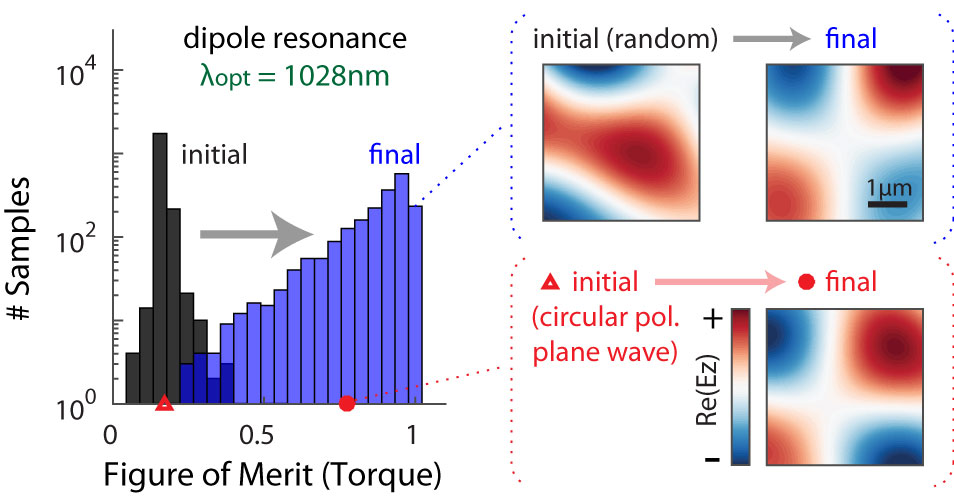}\caption{Distribution of FOM for 2000 randomly chosen incident field configurations
at 1028nm, before (black) and after (blue) optimization. Red marks
on the $x$-axis indicate the initial (triangle) and final (circle)
FOM when the optimization starts from a circularly polarized plane
wave. Representative incident fields are plotted in the right. \label{fig:Distribution-dip}}
\end{figure}
\begin{figure}
\centering{}\includegraphics[height=7cm]{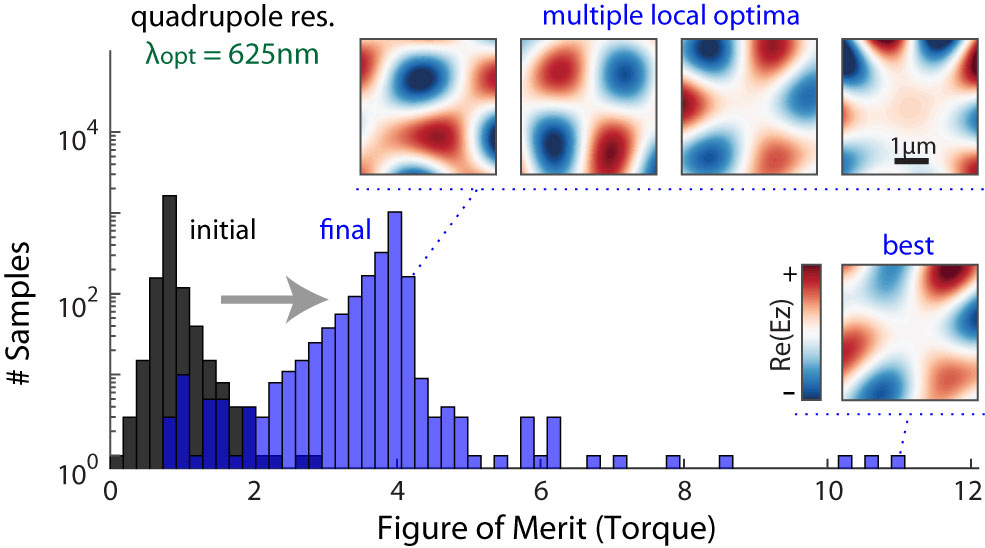}\caption{Distribution of FOM for 2000 randomly chosen incident field configurations
at 625nm. Top inset shows four different field patterns with a near-identical
FOM near the median, and the bottom inset shows the field with the
best FOM. \label{fig:Distribution-quad}}
\end{figure}
The illumination-field design space is nonconvex and littered with
local optima, due primarily to wave-optical interference effects.
We survey this broader design space by restarting our local-optimization
algorithm 2000 times with randomly selected initial configurations
that are constructed using Eq.(\ref{eq:Eexpand}), where the complex
coefficients $a_{i}$ and $b_{i}$ are uniform random numbers bounded
by $\vert a_{i}\vert,\vert b_{i}\vert\leq1$. The results are summarized
in Figs.\ref{fig:Distribution-dip}-\ref{fig:Distribution-quad}
at $\lambda_{\text{opt}}=1028\text{nm}$ (dipole resonance) and 625nm
(quadrupole resonance), respectively. At both wavelengths, we find
that more than 50\% of local optimizations from random starting points
can achieve over 5x enhancement of $\text{FOM}_{T}$ compared to CP
planewave reference, and that the optimized field patterns contain
various combinations of Bessel-basis modes without a systematic convergence
to one over the others. The distributions are plotted in log-scale
to increase the visibility of small bins. 

In Fig.\ref{fig:Distribution-dip}, the median $\text{FOM}_{T}$
at 0.913 is very close to the best $\text{FOM}_{T}$ at 1.01 and the
distribution is predominantly concentrated to the right: the 4 rightmost
bars represent 69\% of all samples, which all acheive over 5x enhancement
of $\text{FOM}_{T}$ compared to CP planewave reference at 0.169 (marked
with a red triangle). The insets show the optimization results from
two different starting points – random field (top) and CP planewave
(bottom) – that reach similar optimized field patterns. We also observe
that a variety of other patterns, dominated by different combinations
of Bessel-basis modes, can produce a nearly identical or superior
FOM.

In Fig.\ref{fig:Distribution-quad}, the final $\text{FOM}_{T}$
distribution is more dispersed between the median at 3.934 and the
best at 10.94, which respectively achieve over 5x and 14x enhancement
compared to CP planewave reference at 0.779. As in Fig.\ref{fig:Distribution-dip},
the results concentrate heavily around the median; however, in Fig.\ref{fig:Distribution-quad} a small number of samples achieve a remarkable
improvement above 14-fold. The top inset shows four different field
patterns that produce a nearly identical $\text{FOM}_{T}$ above the
median, and the bottom inset shows the field-pattern with the highest
$\text{FOM}_{T}$. A comparison of all optimized field patterns at
1028nm and 625nm shows that the latter contains more higher-order
Bessel-basis contributions.

\subsection{Dependence on Illumination Wavelength\label{subsec:Force}}

\begin{figure}
\centering{}\includegraphics[height=6cm]{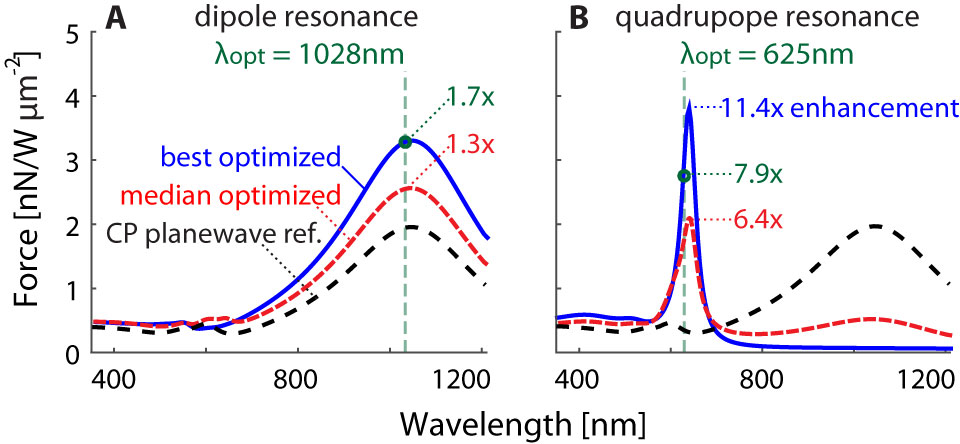}\caption{Force spectrum for two different target wavelengths; (A) $1028\text{nm}$
and (B) $625\text{nm}$. The spectrums for the best (blue line) and
the median (red dashed line) optimized field configurations are each
labeled with the factor of enhancement, with respect to CP planewave
reference (black dashed line). \label{fig:Force-Spectrum} }
\end{figure}
\begin{figure}
\centering{}\includegraphics[height=8cm]{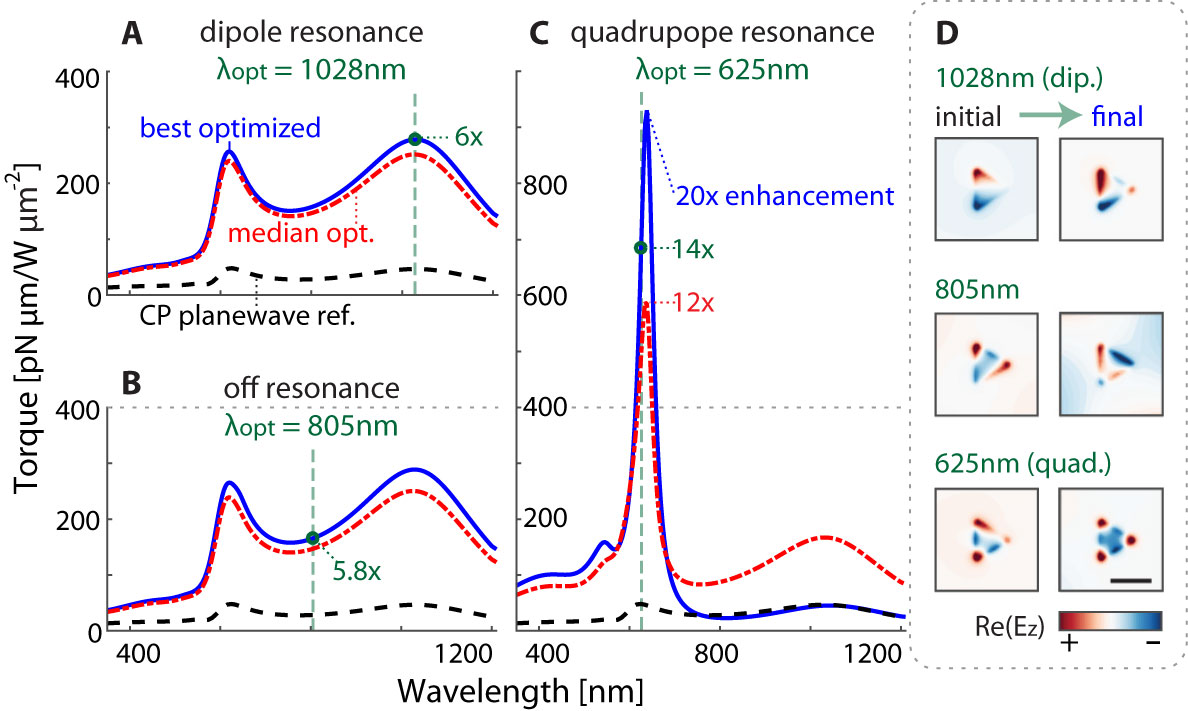}\caption{Torque spectrum for three different target wavelengths; (A) $1028\text{nm}$,
(B) $805\text{nm}$, and (C) $625\text{nm}$. The spectrums for the
best (blue line) and the median (red dashed line) optimized field
configurations are each labeled with the factor of enhancement, with
respect to CP planewave referrence (black dashed line). (D) The total-field
distributions of the initial random field (left) and the final optimized
field (right) for the best optimized field configurations at the three
target wavelengths. Scalebar is $400\text{nm}$. \label{fig:Torque-Spectrum}}
\end{figure}
We further investigate the influence of $\lambda_{\text{opt}}$ by
plotting optical force and torque per incident-field intensity $I_{\text{avg}}$
as a function of illumination wavelength. In Figs.\ref{fig:Force-Spectrum}A-\ref{fig:Force-Spectrum}B,
the reference planewave force spectrum (black dashed line) is identical
in both plots, clearly dominated by a broad dipole resonance with
smaller peaks at higher-order resonances. Through illumination-field
optimization, we can enhance the force at the dipole mode while suppressing
higher-order modes (Fig.\ref{fig:Force-Spectrum}A) and also enhance
the force at the quadrupole mode while suppressing the dipole mode
(Fig.\ref{fig:Force-Spectrum}B). 

In Figs.\ref{fig:Torque-Spectrum}A-\ref{fig:Torque-Spectrum}C,
the reference planewave torque spectrum (black dashed line) is identical
in all three plots and peaks at both dipole and quadrupole resonances
with nearly equal heights (explained in detail in \cite{lee_optical_2014}).
Illumination-field optimization at dipole resonance and off-resonance
achieve a similar 6x-boost at the chosen $\lambda_{\text{opt}}$ value
without suppressing the quadrupole resonance. The optimized total
fields in Fig.\ref{fig:Torque-Spectrum}D at 1028nm and 805nm both
exhibit a $4\pi$ phase change around the circumference of the particle,
resembling a quadrupole resonance.

In \ref{fig:Torque-Spectrum}C, on the other hand, the best optimization
at quadrupole resonance achieves a remarkable 20x improvement while
suppressing much of the dipole resonance; the median optimization
also achieves 12x improvement while suppressing the dipole resonance
to a lesser extent. The optimized total field in Fig.\ref{fig:Torque-Spectrum}D
at 625nm shows a distinct, highly resonant distribution.

When comparing interactions with two different resonances, we were
able to distinguish the impact of the change in resonant field pattern
from the change in the resonance lifetime. Controlling for the change
in lifetime, we found that torque seems to favor higher-order (\emph{e.g.}
quadrupole) resonances with greater angular momentum, while force
($F_{z}$), which closely correlates with extinction power, seems
to favor lower-order (\emph{e.g.} dipole) resonance with greater field
intensity within the particle. With the use of structured illumination,
higher-order resonances can be excited more effectively, which contributes
to higher optical torque after optimization.

Our numerical optimization framework allows a systematic search of
the illumination-field design space to maximize force and torque on
lossy, non-spherical particles with multipolar scattering channels.
In addition, we think a rigorous analytical study of the fundamental
upper bounds on opto-mechanical responses, similar to the analysis
performed on light extinction \cite{miller_fundamental_2014,miller_fundamental_2016},
would be useful in the future. 

\subsection{Robustness of Optimization \label{subsec:Robustness}}

\begin{figure}
\centering{}\includegraphics[height=6cm]{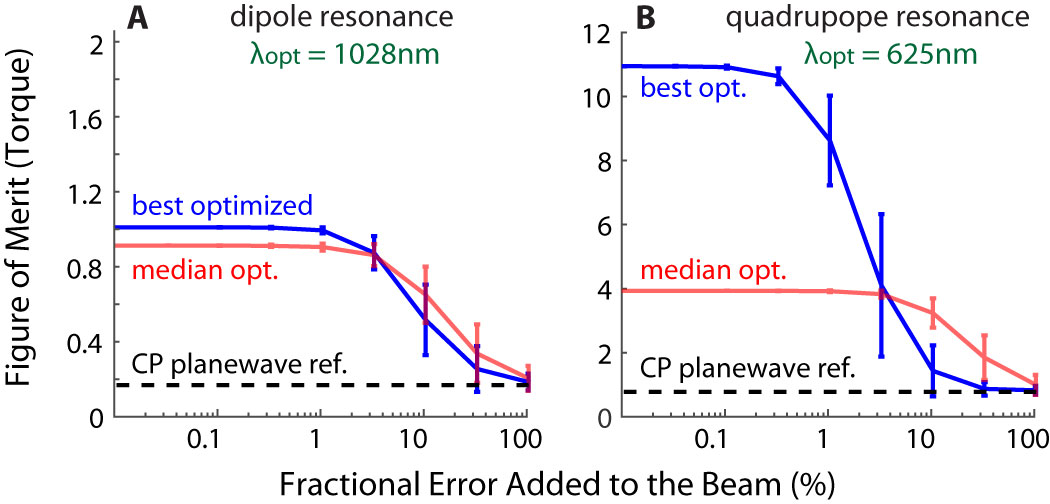}\caption{Robustness of the optimized incident fields, quantified by the decrease
in $\text{FOM}_{T}$ with respect to fractional random error added
to the best (blue) and the median (red) optimized fields, at $\lambda_{\text{dip}}$
(right) and $\lambda_{\text{quad}}$ (left). The error bar represents
the standard deviation for 100 samples.\label{fig:Robustness}}
\end{figure}
Experimental generation of the designed illumination via SLMs may
suffer various types of manufacturing errors \cite{jesacher_wavefront_2007}.
Fig.\ref{fig:Robustness} shows the tradeoff between enhancement
and robustness to experimental errors. The fractional error is added
to the beam using
\begin{equation}
\mathbf{E}_{w}=\sum_{i=0}^{m}(a_{i}+\delta_{ai})\mathbf{M}_{i}+(b_{i}+\delta_{bi})\mathbf{N}_{i},\quad\left(\vert\delta\vert\leq w\cdot\vert a,b\vert_{\infty}\right),
\end{equation}
where $\delta$ is a complex-valued random error bounded by the magnitude
of the largest coefficient multiplied by the fractional weight $0\leq w\leq1$.
Fig.\ref{fig:Robustness} shows the decrease in $\text{FOM}_{T}$
as a function of $w$. At $\lambda_{\text{dip}}$, increasing $w$
from $1\%$ to $10\%$ decreased the best FOM from $0.994$ to $0.517$,
and the median FOM followed a similar trend. On the other hand, $\text{FOM}_{T}$
of the best optimized field at $\lambda_{\text{quad}}$ dropped from
$8.624$ to $1.433$, and the median optimized field changed from
$3.92$ to $3.237$. Fig.\ref{fig:Robustness} demonstrates a clear
tradeoff between field enhancement and error tolerance, as one might
expect due to the need to couple strongly to the underlying particle
resonances. For the tolerance requirements of a given experimental
setup, the approach we outline here could easily be adapted to a robust-optimization
framework \cite{boyd_convex_2004,mutapcic_robust_2009} in which the
expected variability is included and optimized against.

\section{Conclusion}

We present a numerical framework for computer optimization of structured
illumination that maximizes optical force and torque on arbitrary
scatterers, and show a 20-fold enhancement in optical torque per intensity
on an example plasmonic nanoparticle, compared to a circularly polarized
planewave. Previously, the major bottleneck has been the cumbersome
computation. We overcome this bottleneck with a compact cylindrical
Bessel basis and a fast boundary element method. We are optimistic
that such computational framework for 3d vector fields can be generalized
and applied to other design problems in opto-mechanics, nanophotonics,
and 3d imaging.

\section*{Acknowledgement}

The authors thank George Barbastathis for helpful discussions. 



\section*{References}
\begin{thebibliography}{10}
\newcommand{\enquote}[1]{``#1''}

\bibitem{bertsekas_nonlinear_1999}
D.~P. Bertsekas, \emph{Nonlinear programming} (Athena scientific Belmont,
  1999).

\bibitem{powell_fast_1978}
M.~J. Powell, \enquote{A fast algorithm for nonlinearly constrained
  optimization calculations,} in \enquote{Numerical analysis,}  (Springer,
  1978), pp. 144--157.

\bibitem{johnson_nlopt_2014}
S.~G. Johnson, \emph{The {NLopt} nonlinear-optimization package} (2014).

\bibitem{grier_revolution_2003}
D.~G. Grier, \enquote{A revolution in optical manipulation,} Nature
  \textbf{424}, 810--816 (2003).

\bibitem{dholakia_optical_2008}
K.~Dholakia, P.~Reece, and M.~Gu, \enquote{Optical micromanipulation,} Chem.
  Soc. Rev. \textbf{37}, 42--55 (2008).

\bibitem{agarwal_manipulation_2005}
R.~Agarwal, K.~Ladavac, Y.~Roichman, G.~Yu, C.~M. Lieber, and D.~G. Grier,
  \enquote{Manipulation and assembly of nanowires with holographic optical
  traps,} Opt. Express \textbf{13}, 8906--8912 (2005).

\bibitem{kelly_optical_2003}
K.~L. Kelly, E.~Coronado, L.~L. Zhao, and G.~C. Schatz, \enquote{The optical
  properties of metal nanoparticles: the influence of size, shape, and
  dielectric environment,} The Journal of Physical Chemistry B \textbf{107},
  668--677 (2003).

\bibitem{xia_shape-controlled_2005}
Y.~Xia and N.~J. Halas, \enquote{Shape-controlled synthesis and surface
  plasmonic properties of metallic nanostructures,} MRS bulletin \textbf{30},
  338--348 (2005).

\bibitem{heckenberg_generation_1992}
N.~R. Heckenberg, R.~McDuff, C.~P. Smith, and A.~G. White, \enquote{Generation
  of optical phase singularities by computer-generated holograms,} Opt. Lett.
  \textbf{17}, 221--223 (1992).

\bibitem{curtis_dynamic_2002}
J.~E. Curtis, B.~A. Koss, and D.~G. Grier, \enquote{Dynamic holographic optical
  tweezers,} Optics Communications \textbf{207}, 169--175 (2002).

\bibitem{di_leonardo_computer_2007}
R.~Di~Leonardo, F.~Ianni, and G.~Ruocco, \enquote{Computer generation of
  optimal holograms for optical trap arrays,} Optics Express \textbf{15},
  1913--1922 (2007).

\bibitem{chen_generation_2011}
H.~Chen, J.~Hao, B.-F. Zhang, J.~Xu, J.~Ding, and H.-T. Wang,
  \enquote{Generation of vector beam with space-variant distribution of both
  polarization and phase,} Optics letters \textbf{36}, 3179--3181 (2011).

\bibitem{karimi_efficient_2009}
E.~Karimi, B.~Piccirillo, E.~Nagali, L.~Marrucci, and E.~Santamato,
  \enquote{Efficient generation and sorting of orbital angular momentum
  eigenmodes of light by thermally tuned q-plates,} Applied Physics Letters
  \textbf{94}, 231124 (2009).

\bibitem{dolev_surface-plasmon_2012}
I.~Dolev, I.~Epstein, and A.~Arie, \enquote{Surface-plasmon holographic beam
  shaping,} Physical review letters \textbf{109}, 203903 (2012).

\bibitem{schulz_integrated_2013}
S.~A. Schulz, T.~Machula, E.~Karimi, and R.~W. Boyd, \enquote{Integrated multi
  vector vortex beam generator,} Opt. Express \textbf{21}, 16130--16141 (2013).

\bibitem{chen_creating_2015}
C.-F. Chen, C.-T. Ku, Y.-H. Tai, P.-K. Wei, H.-N. Lin, and C.-B. Huang,
  \enquote{Creating {Optical} {Near}-{Field} {Orbital} {Angular} {Momentum} in
  a {Gold} {Metasurface},} Nano Letters \textbf{15}, 2746--2750 (2015).

\bibitem{liu_radiation_2005}
M.~Liu, N.~Ji, Z.~Lin, and S.~Chui, \enquote{Radiation torque on a birefringent
  sphere caused by an electromagnetic wave,} Physical Review E \textbf{72}
  (2005).

\bibitem{liu_light-driven_2010}
M.~Liu, T.~Zentgraf, Y.~Liu, G.~Bartal, and X.~Zhang, \enquote{Light-driven
  nanoscale plasmonic motors,} Nature nanotechnology \textbf{5}, 570--573
  (2010).

\bibitem{lehmuskero_ultrafast_2013}
A.~Lehmuskero, R.~Ogier, T.~Gschneidtner, P.~Johansson, and M.~Käll,
  \enquote{Ultrafast {Spinning} of {Gold} {Nanoparticles} in {Water} {Using}
  {Circularly} {Polarized} {Light},} Nano Letters p. 130624122754005 (2013).

\bibitem{arita_laser-induced_2013}
Y.~Arita, M.~Mazilu, and K.~Dholakia, \enquote{Laser-induced rotation and
  cooling of a trapped microgyroscope in vacuum,} Nature Communications
  \textbf{4} (2013).

\bibitem{chen_optical_2016}
J.~Chen, N.~Wang, L.~Cui, X.~Li, Z.~Lin, and J.~Ng, \enquote{Optical {Twist}
  {Induced} by {Plasmonic} {Resonance},} Scientific Reports \textbf{6}, 27927
  (2016).

\bibitem{novitsky_single_2011}
A.~Novitsky, C.-W. Qiu, and H.~Wang, \enquote{Single {Gradientless} {Light}
  {Beam} {Drags} {Particles} as {Tractor} {Beams},} Physical Review Letters
  \textbf{107} (2011).

\bibitem{sukhov_negative_2011}
S.~Sukhov and A.~Dogariu, \enquote{Negative {Nonconservative} {Forces}:
  {Optical} “{Tractor} {Beams}” for {Arbitrary} {Objects},} Phys. Rev.
  Lett. \textbf{107}, 203602 (2011).

\bibitem{simpson_mechanical_1997}
N.~Simpson, K.~Dholakia, L.~Allen, and M.~Padgett, \enquote{Mechanical
  equivalence of spin and orbital angular momentum of light: an optical
  spanner,} Optics Letters \textbf{22}, 52--54 (1997).

\bibitem{torres_twisted_2011}
J.~P. Torres and L.~Torner, \emph{Twisted {Photons}: {Applications} of {Light}
  with {Orbital} {Angular} {Momentum}} (Wiley New York, 2011).

\bibitem{chen_negative_2014}
J.~Chen, J.~Ng, K.~Ding, K.~H. Fung, Z.~Lin, and C.~T. Chan, \enquote{Negative
  {Optical} {Torque},} arXiv:1402.0621v1  (2014).

\bibitem{lehmuskero_plasmonic_2014}
A.~Lehmuskero, Y.~Li, P.~Johansson, and M.~Käll, \enquote{Plasmonic particles
  set into fast orbital motion by an optical vortex beam,} Optics Express
  \textbf{22}, 4349 (2014).

\bibitem{singer_three-dimensional_2000}
W.~Singer, S.~Bernet, N.~Hecker, and M.~Ritsch-Marte,
  \enquote{Three-dimensional force calibration of optical tweezers,} Journal of
  Modern Optics \textbf{47}, 2921--2931 (2000).

\bibitem{gersborg_maximizing_2011}
A.~R. Gersborg and O.~Sigmund, \enquote{Maximizing opto-mechanical interaction
  using topology optimization,} International Journal for Numerical Methods in
  Engineering \textbf{87}, 822--843 (2011).

\bibitem{hajizadeh_optimized_2010}
F.~Hajizadeh and S.~N. S~Reihani, \enquote{Optimized optical trapping of gold
  nanoparticles,} Optics express \textbf{18}, 551--559 (2010).

\bibitem{tolic-norrelykke_matlab_2004}
I.~M. Tolić-Nørrelykke, K.~Berg-Sørensen, and H.~Flyvbjerg,
  \enquote{{MatLab} program for precision calibration of optical tweezers,}
  Computer Physics Communications \textbf{159}, 225--240 (2004).

\bibitem{polin_optimized_2005}
M.~Polin, K.~Ladavac, S.-H. Lee, Y.~Roichman, and D.~Grier, \enquote{Optimized
  holographic optical traps,} Optics Express \textbf{13}, 5831--5845 (2005).

\bibitem{martin-badosa_design_2007}
E.~Martín-Badosa, M.~Montes-Usategui, A.~Carnicer, J.~Andilla,
  E.~Pleguezuelos, and I.~Juvells, \enquote{Design strategies for optimizing
  holographic optical tweezers set-ups,} Journal of Optics A: Pure and Applied
  Optics \textbf{9}, S267 (2007).

\bibitem{bianchi_real-time_2010}
S.~Bianchi and R.~Di~Leonardo, \enquote{Real-time optical micro-manipulation
  using optimized holograms generated on the {GPU},} Computer Physics
  Communications \textbf{181}, 1444--1448 (2010).

\bibitem{cizmar_holographic_2010}
T.~Čižmár, O.~Brzobohaty, K.~Dholakia, and P.~Zemánek, \enquote{The
  holographic optical micro-manipulation system based on counter-propagating
  beams,} Laser Physics Letters \textbf{8}, 50 (2010).

\bibitem{tao_tao_3d_2011}
T.~T. Tao~Tao, J.~L. Jing~Li, Q.~L. Qian~Long, and X.~W. Xiaoping~Wu,
  \enquote{3d trapping and manipulation of micro-particles using holographic
  optical tweezers with optimized computer-generated holograms,} Chinese Optics
  Letters \textbf{9}, 120010--120013 (2011).

\bibitem{lapointe_towards_2011}
C.~P. Lapointe, T.~G. Mason, and I.~I. Smalyukh, \enquote{Towards total
  photonic control of complex-shaped colloids by vortex beams,} Optics express
  \textbf{19}, 18182--18189 (2011).

\bibitem{j.d._jackson_classical_1962}
{J.D. Jackson}, \emph{Classical {Electrodynamics}}, vol.~3 (Wiley New York,
  1962), third edition ed.

\bibitem{bruning_multiple_1971}
J.~H. Bruning and Y.~T. Lo, \enquote{Multiple scattering of {EM} waves by
  spheres part {I}–{Multipole} expansion and ray-optical solutions,} Antennas
  and Propagation, IEEE Transactions on \textbf{19}, 378--390 (1971).

\bibitem{bohren_absorption_2004}
C.~F. Bohren and D.~R. Huffman, \emph{Absorption and scattering of light by
  small particles} (Wiley-VCH Verlag GmbH \& Co. KGaA, Weinheim, 2004).

\bibitem{stratton_electromagnetic_2007}
J.~A. Stratton, \emph{Electromagnetic theory} (John Wiley \& Sons, 2007).

\bibitem{zhan_cylindrical_2009}
Q.~Zhan, \enquote{Cylindrical vector beams: from mathematical concepts to
  applications,} Advances in Optics and Photonics \textbf{1}, 1 (2009).

\bibitem{rosen_pseudo-nondiffracting_1995}
J.~Rosen, B.~Salik, and A.~Yariv, \enquote{Pseudo-nondiffracting beams
  generated by radial harmonic functions,} JOSA A \textbf{12}, 2446--2457
  (1995).

\bibitem{volke-sepulveda_orbital_2002}
K.~Volke-Sepulveda, V.~Garcés-Chávez, S.~Chávez-Cerda, J.~Arlt, and
  K.~Dholakia, \enquote{Orbital angular momentum of a high-order {Bessel} light
  beam,} Journal of Optics B: Quantum and Semiclassical Optics \textbf{4}, S82
  (2002).

\bibitem{chew_integral_2008}
W.~C. Chew, M.~S. Tong, and B.~Hu, \enquote{Integral equation methods for
  electromagnetic and elastic waves,} Synthesis Lectures on Computational
  Electromagnetics \textbf{3}, 1--241 (2008).

\bibitem{harrington_field_1996}
R.~F. Harrington and J.~L. Harrington, \emph{Field computation by moment
  methods} (Oxford University Press, 1996).

\bibitem{reid_efficient_2013}
M.~T. Reid and S.~G. Johnson, \enquote{Efficient {Computation} of {Power},
  {Force}, and {Torque} in {BEM} {Scattering} {Calculations},} arXiv preprint
  arXiv:1307.2966  (2013).

\bibitem{hamam_coupled-mode_2007}
R.~E. Hamam, A.~Karalis, J.~Joannopoulos, and M.~Soljačić,
  \enquote{Coupled-mode theory for general free-space resonant scattering of
  waves,} Physical review A \textbf{75}, 053801 (2007).

\bibitem{kwon_optimal_2009}
D.-H. Kwon and D.~M. Pozar, \enquote{Optimal characteristics of an arbitrary
  receive antenna,} IEEE Transactions on Antennas and Propagation \textbf{57},
  3720--3727 (2009).

\bibitem{liberal_least_2014}
I.~Liberal, Y.~Ra'di, R.~Gonzalo, I.~Ederra, S.~A. Tretyakov, and R.~W.
  Ziolkowski, \enquote{Least {Upper} {Bounds} of the {Powers} {Extracted} and
  {Scattered} by {Bi}-anisotropic {Particles},} IEEE Transactions on Antennas
  and Propagation \textbf{62}, 4726--4735 (2014).

\bibitem{lee_optical_2014}
Y.~E. Lee, K.~H. Fung, D.~Jin, and N.~X. Fang, \enquote{Optical torque from
  enhanced scattering by multipolar plasmonic resonance,} Nanophotonics
  \textbf{3}, 343--440 (2014).

\bibitem{marston_radiation_1984}
P.~L. Marston and J.~H. Crichton, \enquote{Radiation torque on a sphere caused
  by a circularly-polarized electromagnetic wave,} Physical Review A
  \textbf{30}, 2508 (1984).

\bibitem{friese_optical_1998}
M.~Friese, T.~Nieminen, N.~Heckenberg, and H.~Rubinsztein-Dunlop,
  \enquote{Optical torque controlled by elliptical polarization,} Optics
  letters \textbf{23}, 1--3 (1998).

\bibitem{miller_fundamental_2014}
O.~Miller, C.~Hsu, M.~Reid, W.~Qiu, B.~DeLacy, J.~Joannopoulos, M.~Soljačić,
  and S.~Johnson, \enquote{Fundamental {Limits} to {Extinction} by {Metallic}
  {Nanoparticles},} Physical Review Letters \textbf{112} (2014).

\bibitem{miller_fundamental_2016}
O.~D. Miller, A.~G. Polimeridis, M.~T. Homer~Reid, C.~W. Hsu, B.~G. DeLacy,
  J.~D. Joannopoulos, M.~Soljačić, and S.~G. Johnson, \enquote{Fundamental
  limits to optical response in absorptive systems,} Optics Express
  \textbf{24}, 3329 (2016).

\bibitem{jesacher_wavefront_2007}
A.~Jesacher, A.~Schwaighofer, S.~Fürhapter, C.~Maurer, S.~Bernet, and
  M.~Ritsch-Marte, \enquote{Wavefront correction of spatial light modulators
  using an optical vortex image,} Opt. Express \textbf{15}, 5801--5808 (2007).

\bibitem{boyd_convex_2004}
S.~Boyd and L.~Vandenberghe, \emph{Convex optimization} (Cambridge university
  press, 2004).

\bibitem{mutapcic_robust_2009}
A.~Mutapcic, S.~Boyd, A.~Farjadpour, S.~G. Johnson, and Y.~Avniel,
  \enquote{Robust design of slow-light tapers in periodic waveguides,}
  Engineering Optimization \textbf{41}, 365--384 (2009).

\end{thebibliography}
\end{document}